\documentclass[entropy,article,accept,pdftex,moreauthors]{Definitions/mdpi} 

\firstpage{1} 
\pubvolume{1}
\issuenum{1}
\articlenumber{0}
\pubyear{2022}
\copyrightyear{2022}
\datereceived{} 
\dateaccepted{} 
\datepublished{} 
\hreflink{https://doi.org/} 

\usepackage{amssymb,amsmath,mathrsfs,stmaryrd,mathtools,amsfonts,graphicx,url,booktabs,nicefrac,comment,microtype,tikz-cd,lingmacros,nccmath,tree-dvips, bbm, bm, etoolbox,algorithm,caption,subcaption,color,enumitem, physics, lineno, setspace,multirow}

\Title{Collective dynamics, diversification and optimal portfolio construction for cryptocurrencies}

\TitleCitation{Collective dynamics, diversification and optimal portfolio construction for cryptocurrencies}

\Author{Nick James $^{1,\ddagger}$*\orcidA{}, Max Menzies $^{2,\ddagger}$\orcidB{}}

\AuthorNames{Nick James and Max Menzies}

\AuthorCitation{James, N.; Menzies, M.}

\address{%
$^{1}$ \quad School of Mathematics and Statistics, University of Melbourne, Victoria 3010, Australia; nick.james@unimelb.edu.au \\
$^{2}$ \quad Beijing Institute of Mathematical Sciences and Applications, 
Tsinghua University, Beijing 101408, China; max.menzies@alumni.harvard.edu
}

\corres{Correspondence: nick.james@unimelb.edu.au (N.J.)}

\secondnote{These authors contributed equally to this work.}

\abstract{
Since its conception, the cryptocurrency market has been frequently described as an immature market, characterized by significant swings in volatility and occasionally described as lacking rhyme or reason. There has been great speculation as to what role it plays in a diversified portfolio. For instance, is cryptocurrency exposure an inflationary hedge or a speculative investment that follows broad market sentiment with amplified beta? We have recently explored similar questions with a clear focus on the equity market. There, our research revealed several noteworthy dynamics such as: an increase in the market's collective strength and uniformity during crises, greater diversification benefits across equity sectors (rather than within them), and the existence of a "best value" portfolio of equities. In essence, we can now contrast any potential signatures of maturity we identify in the cryptocurrency market and contrast these with the substantially larger, older and better established equity market. This paper aims to investigate whether the cryptocurrency market has recently exhibited similar mathematical properties as the equity market. Instead of relying on traditional portfolio theory, which is grounded in the financial dynamics of equity securities, we adjust our experimental focus to capture the presumed behavioral purchasing patterns of retail cryptocurrency investors. Our focus is on collective dynamics and portfolio diversification in the cryptocurrency market, and examining whether previously established results in the equity market hold in the cryptocurrency market, and to what extent. Results reveal nuanced signatures of maturity related to the equity market, including the fact that correlations collectively spike around exchange collapses, and identify an ideal portfolio size and spread across different groups of cryptocurrencies.
} 

\begin{document}


\section{Introduction}
\label{Introduction}

One of the topics at the heart of complex systems analysis is the study of financial markets. Financial markets have a diverse range of participants ranging from extremely sophisticated investors leveraging a technological and information advantage to retail investors who may purchase securities based on other fundamental intuitions. One asset class that has seen a significant degree of variance in the sophistication of the investor base is the cryptocurrency market. Over the last few years, the cryptocurrency market has gathered meaningful interest from institutional and retail investors alike. Despite exhibiting tumultuous changes in aggregate assets under management, the overall market has produced substantial growth in total assets since its inception. Given the relative immaturity of the cryptocurrency market, it is important to study the underlying dynamics of the market and contrast optimal trading and portfolio management strategies with that of more traditional asset classes such as the equity market. The main motivation of this paper is to investigate the next stage of the cryptocurrency market's evolution. Although the cryptocurrency market is young, we feel that it may be coming of age and exhibiting signs of maturity, becoming more like the equity market. To assess this, we tactically assess whether certain phenomena such as collective movement, uniformity and diversification benefits are similar to that of the equity market.

It is worth commenting more broadly on financial market dynamics and the wealth of work that has been done on that topic, before we focus on the cryptocurrency market most specifically. There are a variety of academic communities who have studied financial market dynamics and evolutionary changes in structural dynamics such as those in applied mathematics, complex systems and econometrics \citep{Fenn2011,Laloux1999,Mnnix2012}. A wide range of data scientific methodologies have been used to study evolutionary dynamics in financial assets such as linear algebraic-inspired techniques \cite{Laloux1999,Kim2005,Pan2007,Wilcox2007}, spectral methods such as random matrix theory \cite{Conlon2007,Bouchaud2003,Burda2004,Sharifi2004,Fenn2011}, a variety of unsupervised learning methodologies \cite{Heckens2020,Jamesfincovid}, change point detection \cite{James2021_crypto,JamescryptoEq} and a litany of statistical modeling techniques \cite{Ferreira2020}.

Another topic of substantial interest to the financial markets community is that of nonstationarity, regime switching and the time-varying nature of model parameters for phenomena such as volatility. Such research dates back to autoregressive conditionally heteroskedastic (ARCH) models \cite{Engle1982}, generalized ARCH (GARCH) \cite{Bollerslev1986} and stochastic adaptations such as stochastic volatility models \cite{Taylor1982,Taylor1986,Taylor1994}. Recently, many researchers have explored adaptions to these fundamental models explicitly capturing dynamics exhibited by various time series. Some of these models include exponential general autoregressive conditional heteroskedastic models \cite{Nelson1991}, Glosten-Jagannathan-Runkle-GARCH \cite{GLOSTEN1993}, Treshold-GARCH \cite{Zakoian1994} and T-SV \cite{So2002}, Markov switching GARCH \cite{Cai1994,Hamilton1994,Gray1996} and MS-SV \cite{Lam1998}. Many financial mathematicians have also adopted Bayesian estimation methodologies \cite{Andersen2003,Hwang2007,Hansen2011,james2021_spectral}, generally citing the need for uncertainty quantification in estimating model parameters. These modeling techniques have been widely applied to the study of several asset classes including equities, cryptocurrencies and fixed income \cite{Cerqueti2020,Wan2017,Stehlk2017,ChiaShangJamesChu1996,Chen2018,Cerqueti2019}. Finally, we would be remiss not to mention the wide range of techniques in time series analysis that have been used to study financial problems \cite{Drod2021_entropy,Liu1997,james2021_portfolio,Basalto2007,Wtorek2021_entropy,james_arjun,Drod2001,james2022_stagflation,Gbarowski2019,james2021_MJW}, including cryptocurrencies \cite{Sigaki2019,Drod2020_entropy,James2021_crypto2,Drod2020,Wtorek2020,Drod2018,Drod2019,DrodKwapie2022_crypto,DrodWtorek2022_crypto,DrodWtorek2023_crypto} and diverse fields in socio- and econophysics \cite{james2021_hydrogen,Perc2020,Machado2020,james2023_hydrogen2,Merritt2014,james2020_Lp,Khan2020,james2022_CO2,Manchein2020,Ribeiro2012,james2021_TVO,Li2021_Matjaz,Blasius2020,James2023_terrorist,Clauset2015,james2022_guns,Perc2013,Singh2022_Watorek,james2021_olympics}

Another topic of great interest across asset classes is the topic of portfolio optimization, and more generally, the essence of portfolio construction. The quantitative finance and econometrics communities have studied core issues related to portfolio diversification, where portfolios are optimized with respect to different objective functions \cite{Markowitz1952,Sharpe1966,Almahdi2017,Calvo2014,Soleimani2009,Vercher2007,Bhansali2007,Moody2001}. More broadly, financial market dynamics are universally difficult to model. The seminal work of Markowitz in 1952 \cite{Markowitz1952} proposed the concept of diversification as a superior framework for investing in multiple securities at a time. The principle underpinning diversification is built upon disassociating the risk of an individual and particular financial asset into a market (systematic) risk component and an asset-specific risk, called unsystematic risk. Diversification essentially equates to smoothing (or averaging over) unsystematic risk by investing in an appropriately large number of individual assets, which leads candidate investment portfolios only exposure being that inherently due to market risk.

In recent work, the authors of this work and collaborators \cite{james_georg} perform a thorough inspection of diversification properties from the perspective of a pure equity portfolio. Precisely, they explore the changing diversification benefit of various portfolios spread across a range of industry sectors. While in more recent years investor composition has broadened to include the likes of quantitative and high-frequency investors, active investment management has historically been dominated by fundamental investors who make investment decisions based on the future potential of companies relative to market valuations (most commonly, the earnings the company produces relative to its share price). The authors hypothesized that there is more substantial diversification benefit investing across sectors, rather than within them. Indeed, different sectors exhibit varying performance during distinct market periods: some sectors may outperform in buoyant equity markets (such as information technology and often energy), while other sectors outperform in declining equity markets (such as healthcare, consumer staples and utilities).

The authors confirmed this hypothesis, producing four primary findings. First, they use time-varying PCA to highlight that the collective behavior of equities spikes during market crises, rendering diversification far less effective. Second, they demonstrate that various community detection algorithms such as modularity are unable to distinguish between heterogeneous equity sector dynamics during times of crisis. By contrast, during more buoyant equity market periods, equity sector behaviors are more easily distinguished. Third, they introduce a new metric to quantify the uniformity of market impact across equity sectors. There, they show substantial variance across the uniformity of market impact across independent equity sectors. Finally, they demonstrate that a best value equity portfolio exists with respect to evolutionary diversification benefits. They show that a portfolio of size 36, where 4 equities are sampled randomly from 9 different equity sectors, provides comparable diversification benefit to a portfolio of size 81, where 9 equities are randomly sampled from 9 equity sectors. Our critical focus is exploring diversification benefits for cost-conscious retail investors. These are investors who are intelligent, may be financially interested but lack the resources to trade frequently in an efficient manner.

With respect to signature of maturity, the cryptocurrency market is very much in its infancy when compared to the equity market. Cryptocurrency sectors are not well defined, and it is often hard to differentiate behaviors between cryptocurrency sectors \cite{JamescryptoEq}. If one explores candidate cryptocurrency sector themes online, categories such as wallet, web3, yield farming, play to earn, energy, decentralized finance, distributed computing and cybersecurity can be found. However, these categories frequently overlap or differ from source to source, and it is not necessarily clear how the behaviors of these cryptocurrency sectors relate to the underlying economy. In fact, it is unclear just how often cryptocurrencies are purchased with the underlying coin sector or thematic within the digital ecosystem in mind. We suspect that this phenomenon is especially pronounced among less sophisticated retail investors - where coins may be bought and sold based on factors such as their recent price and volume movements, and overall macroeconomic trends. Accordingly, in this work we turn to the cryptocurrency market and adapt our experiments to test for alternative diversification strategies among retail cryptocurrency investors. Rather than testing diversification effectiveness among equity sectors, we use tranches of cryptocurrency market capitalizations to proxy sectors. We suspect that many cryptocurrency investors buy securities from platforms where they simply scan a list of assets which are ordered by market capitalization, and are unaware of many coins' association with a deeper role in the digital economy. We feel that this is an original and suitable measure of different ``classes'' of cryptocurrencies.  Here, we apply the same fundamental methodologies to the cryptocurrency market as a means of testing the levels of maturity and sophistication among the cryptocurrency market.

This paper is structured as follows. In Section \ref{sec:data}, we outline the data that we use in this paper. In Section \ref{sec:marketstructure}, we study the evolution of the collective dynamics of the cryptocurrency market. We compare this findings to what has been observed over 20 years in the equity market, and draw conclusions regarding the cryptocurrency market's signatures of maturity. In Section \ref{sec:portfoliosampling}, we turn to the theme of optimal portfolio construction among cryptocurrency portfolios. There we study marginal diversification benefit as additional cryptocurrency sector deciles, and cryptocurrencies within deciles are sequentially added to a portfolio. In Section \ref{sec:conclusion}, we conclude and summarize our findings regarding recent signatures of maturity in the cryptocurrency market.

\section{Data}
\label{sec:data}

Our data are chosen as follows. Our window of analysis ranges from July 1, 2019 to February 14, 2023. As of the final day of our analysis window, we drew the top 75 cryptocurrencies by market capitalization from Yahoo Finance \cite{Yahoo_crypto}, and restricted these to those with a price history dating back to July 1, 2019. This left 42 cryptocurrencies - we then discarded the two smallest, leaving $N=40$ cryptocurrencies and their closing price data over $T=1325$ days. The window of analysis includes periods of major disruption for cryptocurrencies, such as the COVID-19 market crash in 2020, the BitMEX exchange market crash \cite{Forbes_Bitmex}, and the collapse of the FTX exchange in late 2022 \cite{NYT_FTX}. We divide the 40 remaining cryptocurrencies into 10 deciles each with four cryptocurrencies based on market capitalization as of the end of the analysis window. We list all cryptocurrencies analyzed in this paper in Table \ref{tab:cryptolist}.

\begin{table}[h]
\centering
\begin{tabular}{|c|c|c|}
\hline
\textbf{Cryptocurrency} & \textbf{Ticker} & \textbf{Decile} \\
\hline
Bitcoin & BTC & 1 \\
Ethereum & ETH & 1 \\
Tether & USDT & 1 \\
Binance Coin & BNB & 1 \\
USD Coin & USDC & 2 \\
XRP & XRP & 2 \\
Cardano & ADA & 2 \\
Polygon & MATIC & 2 \\
Dogecoin & DOGE & 3 \\
Litecoin & LTC & 3 \\
TRON & TRX & 3 \\
Wrapped Bitcoin & WBTC & 3 \\
Chainlink & LINK & 4 \\
Cosmos & ATOM & 4 \\
UNUS SED LEO & LEO & 4 \\
OKB & OKB & 4 \\
Ethereum Classic & ETC & 5 \\
Filecoin & FIL & 5 \\
Monero & XMR & 5 \\
Bitcoin Cash & BCH & 5 \\
Stellar & XLM & 6 \\
VeChain & VET & 6 \\
Crypto.com Coin & CRO & 6 \\
Algorand & ALGO & 6 \\
Quant & QNT & 7 \\
Fantom & FTM & 7 \\
Tezos & XTZ & 7 \\
Decentraland & MANA & 7 \\
EOS & EOS & 8 \\
Bitcoin BEP2 & BTCB & 8 \\
Theta Network & THETA & 8 \\
TrueUSD & TUSD & 8 \\
Rocket Pool & RPL & 9 \\
Chiliz & CHZ & 9 \\
USDP Stablecoin & USDP & 9 \\
Huobi Token & HT & 9 \\
KuCoin Token & KCS & 10 \\
Bitcoin SV & BSV & 10 \\
Dash & DASH & 10 \\
Zcash & ZEC & 10 \\
\hline
\end{tabular}
\caption{Cryptocurrencies, their tickers, and decile allocations}
\label{tab:cryptolist}
\end{table}

\section{Collective dynamics and uniformity}
\label{sec:marketstructure}

Let $c_i(t)$, $i=1,\dots,N$, $t=0,\dots,T$ denote the multivariate time series of daily closing prices among our collection of $N$ cryptocurrencies. Let $r_i(t)$ be the multivariate time series of log returns $i=1,\dots,N$, $t=1,\dots,T$, defined as 
\begin{align}
\label{eq:logreturns}
r_{i}{(t)} &= \log \left(\frac{c_i{(t)}}{c_i{(t-1)}}\right).
\end{align}

In this section, we analyze correlation matrices of log returns across rolling time windows of length $\tau$; in this paper, we choose $\tau=90$ days. We standardize the log returns over such a window $[t-\tau+1,t]$ by defining ${R}_i(s) = [r_i(s) - \langle r_i \rangle] / \sigma(r_i)$ where $\langle . \rangle $ denotes the average over the time window $[t-\tau+1,t]$ and $\sigma$ the associated standard deviation. Let $\bf{R}$ be the $N \times \tau$ matrix defined by ${R}_{is}={R}_i(s)$ with $i=1,\dots,N$ and $s=t-\tau+1,\dots,t$, and then the correlation matrix $\bm \Psi$ is then defined as 
\begin{align}
\label{eq:corrmatrix}
\bm{\Psi}(t) = \frac{1}{\tau} {\bf{R}} {\bf{R}}^T.
\end{align}
Individual entries describing the correlation between cryptocurrencies $i$ and $j$ can be written as
\begin{align}
\label{eq:rhodefn}
    \Psi_{ij}(t)=\frac{1}{\tau} \frac{\sum_{s=t-\tau+1}^t (r_i(s) - \langle r_i \rangle)(r_j(s) - \langle{r}_j \rangle)}{\left(\sum_{s=t-\tau+1}^t (r_i(s) - \langle r_i \rangle)^2 \right)^{1/2} \left( \sum_{s=t-\tau+1}^t (r_j(s) - \langle r_j \rangle)^2\right)^{1/2}},
\end{align}
for  $1\leq i, j\leq N$. We may analogously define the cross-correlation matrices for each individual decile by restricting $i$ and $j$ to be chosen from a set of indices corresponding to that decile.

\begin{figure}[H]
    \begin{subfigure}[b]{\textwidth}
        \includegraphics[width=\textwidth]{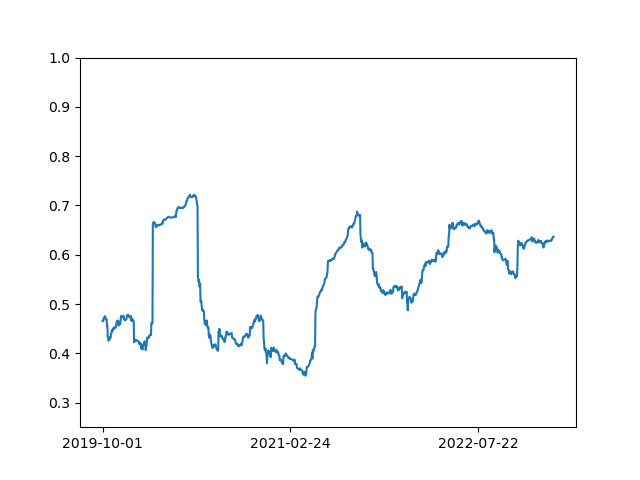}
        \caption{}
        \label{fig:lambda1_all}
    \end{subfigure}
    \begin{subfigure}[b]{\textwidth}
        \includegraphics[width=\textwidth]{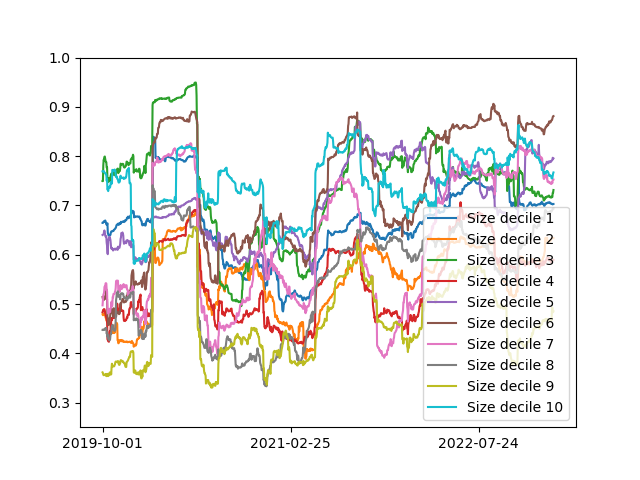}
        \caption{}
        \label{fig:lambda1_deciles}
    \end{subfigure}
    \caption{Normalized leading eigenvalue $\tilde{\lambda}_1(t)$ of the cross-correlation matrix as a function of time, for (a) the entire collection of cryptocurrencies and (b) the ten deciles. Like the equity market, collective correlations spike during market crises, such as COVID-19, and the collapse of exchanges BitMEX and FTX.}
    \label{fig:lambda1}
\end{figure}

All entries $\Psi_{ij}$ lie in the interval $[-1,1]$. $\bm\Psi$ is a symmetric and positive semi-definite matrix with real and non-negative eigenvalues $\lambda_i(t)$, so we order them as $\lambda_1 \geq \cdots \geq \lambda_N \geq 0$. All the diagonal entries of $\bm\Psi$ are equal to 1, so the trace of $\bm\Psi$ is equal to $N$. Thus, we may normalize the eigenvalues by defining by $N$, to wit, $\tilde{\lambda}_i = \frac{\lambda_i}{\sum^{N}_{j=1} \lambda_j}= \frac{\lambda_i}{N}$. We display the rolling normalized first eigenvalue $\tilde{\lambda}_1(t)$ for both the entire collection of cryptocurrencies and the 10 deciles in Figure \ref{fig:lambda1}.

In Figure \ref{fig:lambda1_all}, we see particular periods of heightened collective correlation between cryptocurrencies. In particular, we see extended periods of high correlation in early 2020, coinciding with COVID-19 and the BitMEX crash, and toward the end of 2022, reflecting the tumultuous period around the collapse of FTX. These are perhaps the most significant moments of collective crisis in the cryptocurrency market in the last three years. These broad trends are reflected on a decile-by-decile basis in Figure \ref{fig:lambda1_deciles}, where each individual decile exhibits a rise in collective correlations in these two periods.

To a nuanced extent, this is a signature of growing maturity in the cryptocurrency market. Specifically, crisis periods are observed; there is a fairly robust time differential between crises; collective correlations rise during crises and fall outside these periods; such effect is seen rather uniformly among different sectors of the market. However, we must remark that the extent of maturity does not coincide with more established markets such as the equity market. The time differential between peaks in collective correlations is still notably shorter than it is for equities, for example the large time differential between the Dot-com bubble, the global financial crisis and COVID-19 crash. Moreover, the strength of collective correlations between deciles varies significantly, despite their sharing temporal patterns. Some deciles, such as the third, exhibit significantly higher collective behaviors than others such as the second, fourth, and ninth, whereas these behaviors are much more uniform for equity indices.

Next, we turn to an analysis of the leading \emph{eigenvector} ${\bf v}_1$ to complement our study of the leading eigenvalue. We analyze its \emph{uniformity} via the following computation:
\begin{align}
\label{eq:normalisedip}
h(t) = \frac{|\langle {\bf v}_1,\bm{1}\rangle |}{ \|{\bf v}_1\| \| \bm{1} \|  },
\end{align}
where $\bm{1}=(1,1,\dots,1) \in \mathbb{R}^N$ is a uniform vector of 1's. We may compute this for both the entire collection of cryptocurrencies and individual deciles, analogously to the leading eigenvalue. We observe that $h(t)\le 1$ with $h(t)=1$ if and only if ${\bf v}_1={\bm 1}$ (up to scalar multiplication). In this case, all cryptocurrencies carry the same amount of variance in the ``market effect'' summarized by $\tilde{\lambda}_1(t)$. This can be used to quantify the potential benefit of diversification: increased values of $h(t)$ indicate increased interchangeability of cryptocurrencies in the ``market'' and hence less opportunity for diversification or judicious selection of individual cryptocurrencies.   

\begin{figure}[H]
    \begin{subfigure}[b]{\textwidth}
        \includegraphics[width=\textwidth]{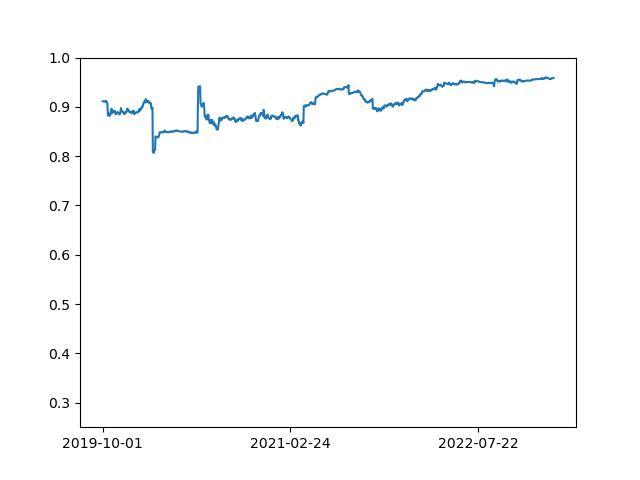}
        \caption{}
        \label{fig:innerproduct_deciles}
    \end{subfigure}
    \begin{subfigure}[b]{\textwidth}
        \includegraphics[width=\textwidth]{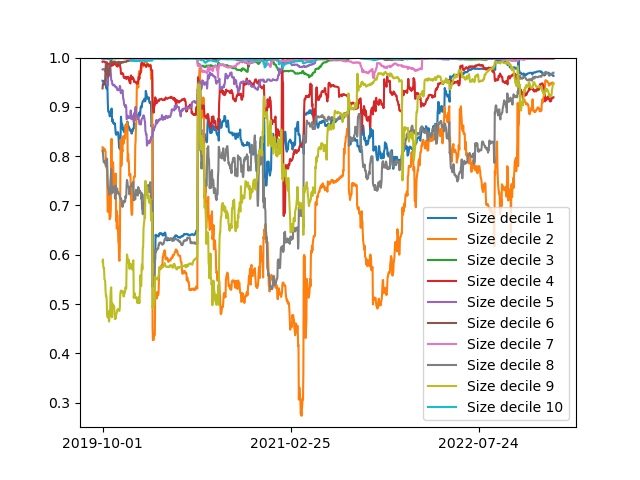}
        \caption{}
        \label{fig:innerproduct_all}
    \end{subfigure}
    \caption{Uniformity $h(t)$ of the leading eigenvector ${\bf v}_1$ of the cross-correlation matrix as a function of time, for (a) the entire collection of cryptocurrencies and (b) the ten deciles. Results are dramatically different compared to the equity market, with numerous deciles exhibiting strikingly low uniformity scores over time.}
    \label{fig:innerproduct}
\end{figure}

We display the rolling uniformity of the first eigenvector $h(t)$ for both the entire collection of cryptocurrencies and the 10 deciles in Figure \ref{fig:innerproduct}. Unlike Figure \ref{fig:lambda1}, we observe a substantial difference compared to the equity market. In the case of the equity market, the uniformity for each sector and the entire market are consistent with the degree of collectivity. The degree of uniformity spikes during market crises such as the dot-com bubble GFC and COVID-19. This spike during market crises occurs for sectors (when studied independently) as well as across the entire market. The cryptocurrency market produces dramatically different findings to that of the equity market. Most notably, we see that there is substantial differences between the uniformity of independent sectors of the cryptocurrency market with that of the equity market. The cryptocurrency market clearly exhibits less uniformity during crises (which we see during the COVID-19 market crisis), and significantly higher variance between sectors of securities throughout our window of analysis. This is the opposite finding to the equity market, where industry sectors exhibited more uniformoity during crises. Another point to note is the stark contrast in how low the $h(t)$ values reach when comparing the two asset classes. In the case of equities, there is a clear lower bound around the value of 0.75, while for cryptocurrencies we see two groups of cryptocurrencies reach values below 0.5 (with one reaching less than 0.3) during our analysis window. Our analysis therefore suggests that we see less persistent and amplified uniformity among cryptocurrencies when compared to equities.

\section{Portfolio sampling}
\label{sec:portfoliosampling}

In this section, we perform an extensive sampling procedure to explore how diversification benefits depend on the number of cryptocurrencies held in a portfolio and on the number of decile sectors from which to choose those cryptocurrencies. Motivated by Section \ref{sec:marketstructure}, we choose the normalized leading eigenvalue $\tilde \lambda_1(t)$ to quantify the potential diversification benefit. We investigate the diversification benefits of portfolios that consist of $mn$ cryptocurrencies such that $n$ cryptocurrencies are drawn from $m$ separate deciles. Both the individual cryptocurrencies and the sector deciles are drawn randomly and independently with uniform probability. We draw $D=500$ portfolios for each ordered pair $(m,n)$.

To quantify the potential diversification benefit for a portfolio consisting of $mn$ cryptocurrencies, we determine the $mn\times mn$ correlation matrix $\bm \Psi$ for each draw and calculate the associated normalized first eigenvalue $\tilde{\lambda}_{m,n}(t)$. Again, we use a rolling time window of length $\tau=90$ days when determining the cross-correlation matrix. In particular, we record the 50th percentile (median) of the $D$ values, which we denote $\tilde{\lambda}_{m,n}^{0.50}(t)$.

We analyze this quantity in two experiments. First, we compute the temporal mean of the median of the normalized first eigenvalue 
\begin{align}
  \mu_{m,n}= \frac{1}{T-\tau+1} \sum_{t=\tau}^T \tilde{\lambda}_{m,n}^{0.50}(t)
\end{align}
as a measure of the diversification benefit of a portfolio with $n$ cryptocurrencies in each of $m$ decile sectors. Table \ref{tab:samplemeans} records these means $\mu_{m,n}$ for cryptocurrency portfolios across values of $(m,n)$ for $1\le m \le  10$ and $1\le n\le 4$.

Table \ref{tab:samplemeans} shows the mean $\mu_{m,n}$ of the median normalized eigenvalue $\tilde{\lambda}^{0.50}_{m,n}(t)$ for various combinations of cryptocurrency sectors, and randomly sampled cryptocurrencies within each sector. We also mark in red a ``greedy path'' to decrease the overall average $\mu_{m,n}$ (that is, increase the overall diversification benefit of a portfolio) by greedily increasing either $m$ or $n$ at each stage. There are several key findings from this analysis. First, the overall structural finding with respect to optimal portfolio construction strongly resembles that of the equity market in \cite{james_georg}. We see incrementally greater benefit in diversifying across sectors rather than within them, and we see significant reduction in marginal diversification benefit once a portfolio reaches a critical mass of securities (sampled from different sectors). This leads to the existence of a ``best value'' cryptocurrency portfolio, like that seen in the equity market. This findings is slightly surprising, and may support our hypothesis, that retail cryptocurrency investors diversify across cryptocurrency market capitalization levels. Indeed, this may occur in the absence of clearly defined sector themes, which may exhibit different performance during different parts of the business cycle. As investors come to better understand cryptocurrencies, and cryptocurrencies related to separate aspects of the digital economy begin to perform differently during various market conditions, this diversification benefit may slightly alter and amplify. That is, rather than cryptocurrency market capitalization being a primary discriminator in diversified performance we may see a closer resemblance to the equity market with cryptocurrency sector themes more closely resembling equity dynamics.

\begin{table*}
\centering
\begin{tabular}{c|rrrr}
  \hline
 & \multicolumn{4}{c}{Number of cryptocurrencies per sector} \\
  \hline
Number of sectors & 1 & 2 & 3 & 4 \\ 
  \hline
    1 & \textcolor{red}{1}  & \textcolor{red}{0.759} & 0.668 & 0.645 \\
  2 & 0.774 & \textcolor{red}{0.651} & \textcolor{red}{0.598} & 0.587  \\ 
 3 & 0.681 & 0.605 & \textcolor{red}{0.581} & 0.576 \\ 
  4 & 0.641 & 0.587 & \textcolor{red}{0.572} & \textcolor{red}{0.565} \\ 
  5 & 0.613 & 0.583 & 0.565 & 0.559 \\ 
  6 & 0.607 & 0.57 & 0.565 & 0.557 \\ 
  7 & 0.593 & 0.565 & 0.559 & 0.555 \\ 
  8 & 0.582 & 0.564 & 0.557 & 0.552 \\ 
  9 & 0.552 & 0.565 & 0.557 & 0.553 \\ 
  10 & 0.581 & 0.560 & 0.554 & 0.552 \\ 
   \hline
\end{tabular}
\caption{Average $\mu_{m,n}$ of the median normalized eigenvalue $\tilde{\lambda}_{m,n}^{0.50}(t)$ for different pairs of $m$ sectors and $n$ cryptocurrencies per sector. In red we display a greedy path that aims to increase the total diversification benefit (by decreasing $\mu_{m,n}$) at each step. We identify a best value cryptocurrency portfolio consisting of 4 sectors and 4 cryptocurrencies per sector. This (4,4) portfolio has nearly the same diversification benefit as the largest possible (10,4) portfolio, as we will also show in the next experiment.}
\label{tab:samplemeans}
\end{table*}

In the second experiment, we investigate which portfolio combinations $(m,n)$ share the most similar evolution in their collective dynamics. For this purpose, we perform hierarchical clustering on the distance metric 
\begin{align}
\label{eq:distanced}
    d((m,n),(m',n'))= \frac{1}{T-\tau+1} \sum_{t=\tau}^T| \tilde{\lambda}_{m,n}^{0.50}(t) - \tilde{\lambda}_{m',n'}^{0.50}(t)|,
\end{align}
which computes the average absolute difference between the median eigenvalues of two portfolios $(m,n)$ and $(m',n')$. This results in a $40 \times 40$ distance matrix for $1\le m \le  10, 1\le n\le 4$. Hierarchical clustering is a convenient and easily visualizable tool to reveal proximity between different elements of a collection. Here, we perform agglomerative hierarchical clustering using the average-linkage criterion \cite{Mllner2013}. The algorithm works in a bottom-up manner, where each ordered pair $(m,n)$ starts in its own cluster, and pairs of clusters are successively merged as one traverses up the hierarchy.

\begin{figure}
    \centering
    \includegraphics[width=\textwidth]{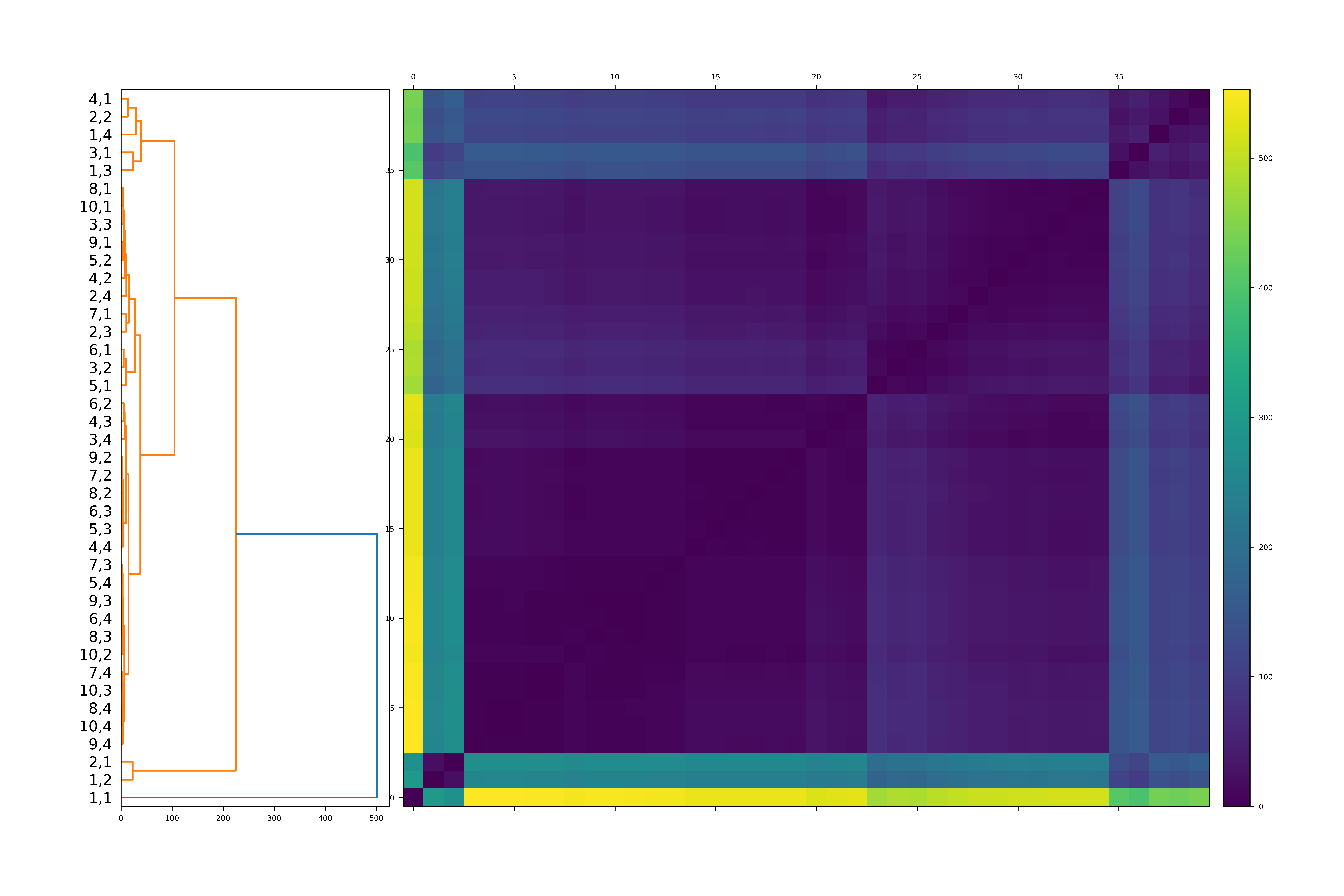}
    \caption{Results of hierarchical clustering applied to (\ref{eq:distanced}) between ordered pairs $(m,n)$. A large majority cluster confirms the finding of Table \ref{tab:samplemeans} that the (4,4) portfolio is closely similar to the full (10,4) portfolio in its diversification benefit over time.}
    \label{fig:m_n_dend}
\end{figure}

The results of hierarchical clustering are displayed in Figure \ref{fig:m_n_dend}. The resulting structure is interesting. The dendrogram consists of four predominant groups of clusters. There is an easily identified outlier cluster, consisting of the smallest portfolios that provide the least diversification benefit. This cluster, located at the bottom of the dendrogram, includes portfolio combinations such as (1,1), (1,2) and (2,1). The second least diversified cluster is located at the top of the dendrogram, and includes portfolio combinations such as (1,3), (1,4) and (4,1). Below this, is a significantly larger cluster consists of portfolio combinations such as (8,1), (3,3) and (4,2). Finally the largest, most well diversified fourth cluster consists of portfolio combinations ranging from (4,3) through to (10,4). The size and range of portfolio combinations within this cluster have interesting implications for risk management in cryptocurrency portfolio diversification. The fact that combination (4,3) is in the same cluster as portfolio (10,4) suggests that comparable risk mitigation can be realized with a portfolio of size 12, when compared to a portfolio of size 40. This finding is not dissimilar to that proposed in \cite{james_georg}, where a ``best value'' portfolio (9,4) is shown to provide comparable diversification benefit to a (9,9) portfolio. Furthermore, the sheer size of this cluster indicates that one may require a lower cardinality portfolio in cryptocurrency portfolio management than in equities when trying to attain a ``best value'' portfolio.

\section{Discussion}
\label{sec:discussion}

This paper has investigated the cryptocurrency market with a focus on collective correlation dynamics and portfolio diversification benefits across different market capitalization size deciles. We choose to investigate deciles as an analogue of industry sectors in the equity market motivated by the vagueness of existing cryptocurrency sectors and the hypothesis that many retail investors use size as a primary means of diversifying across highly different cryptocurrencies. Throughout the paper, we have consistently observed signatures of maturity in the cryptocurrency market with nuanced differences relative to established patterns in the (much) more mature equity market.

In Section \ref{sec:marketstructure}, we analyzed collective dynamics of the cryptocurrency market, focusing on collective correlation strength and market uniformity summarized in the leading eigenvalue and eigenvector of the correlation matrix, examining both the full market and individual deciles. Our first finding is that collective correlations spike during market crises connected to cryptocurrency exchange crashes; this occurs in every decile and closely resembles analogous behavior in the more mature equity market. Other findings of this section portray a more nuanced picture of differences between the cryptocurrency and equity markets. While collective correlations spike across every decile during a crisis, it is not the case that correlations within each decile sector are uniformly higher than collective correlations across the whole market, as we previously observed for the equity market. In addition, the uniformity $h(t)$ of different deciles over time exhibited a finding highly dissimilar from the equity market. This was the most significant difference relative to the equity market observed in this paper. While the uniformity (measuring the uniform of different assets contributing toward the first principle component) was close to maximal 1 for every sector in the equity markets, that finding was not at all observed for the deciles of the cryptocurrency market. Curiously, it was observed for just two deciles consistently over time, but not the others. In addition, uniformity within deciles dropped during market crises, the opposite finding for the equity market.

These findings have significant implications for alpha generation in the cryptocurrency market. The fact that collective correlations are so pronounced during market crises implies that alpha-generating opportunities based on systematic market movements would be more predictable and successful, if done so on the short side. During market crises, correlations translate upward and cryptocurrencies of all sizes tend to decline. This would suggest that fundamentally-driven investment strategies may be more successful when implemented during buoyant equity markets, where there is less correlation among underlying securities. By contrast, during market crises (which are typically coincident among the equity and cryptocurrency asset classes) the collective strength of the market is so strong that the weight of underlying security investments driven by bottom-up analysis may be washed away by the sheer weight of money.

In Section \ref{sec:portfoliosampling}, our portfolio sampling experiment investigated the diversification benefit of portfolios of total size $mn$ spread evenly across $m$ separate deciles. In a greedy experiment, we demonstrated that greater diversification benefit is generally obtained by increasing the number of decile sectors rather than the number of cryptocurrencies per decile, a result analogous to that observed for the cryptocurrency market. We followed this up with a careful experiment clustering different temporal trajectories of median normalized eigenvalue functions $\tilde{\lambda}_{m,n}(t)$. A large majority cluster showed a similar result as observed for the equity market, that a portfolio of spread (4,4) had near-identical diversification benefit as our maximal size (10,4) cryptocurrency collection.

Our findings in this section may drive decision-making for optimal portfolio construction for cryptocurrency investors. First, the emergence of a low-cardinality highly diversified portfolio implies that retail investors may gain exposure to high-quality diversification at low-cost. When contrasting this analysis with that of the equity market, if we were to assume equivalent transaction costs and equivalent periodicity of portfolio rebalancing, the cryptocurrency may be a more retail-friendly market for easy access and portfolio diversification. Of course, given that the equity market is so sophisticated, there is a large number of index-tracking and factor-based investment strategies which may benefit retail investors. This could be an opportunity for asset managers and large investment firms who are looking to create cryptocurrency investment products, and is certainly a signature of the market's maturity. Finally, our analysis supports the notion that the cryptocurrency market may be a suitable environment for skilful stock pickers. We have highlighted that a portfolio of just 16 stocks produces low correlation, and significant diversification benefits. This would indicate that an investment portfolio built upon a smaller number of high-conviction ideas could thrive in the cryptocurrency market.

There are several insights contained within concerning the cryptocurrency market's levels of maturity. First, the overall structure of the aforementioned hierarchical clustering is highly similar to that of the equity market. We have identified heterogeneous clusters of diversification benefit, and highlight the existence of a ``best value'' cryptocurrency portfolio where comparable diversification benefit is attained with relatively fewer securities held in a portfolio. Second, a crucial corollary of this finding is that retail investors with limited ability to hold complex portfolios of many cryptocurrencies may be sufficiently diversified with a relatively small portfolio across just 16 cryptocurrencies. However, there are some key differences to the equity market. First, the link between underlying cryptocurrencies' business functions (at least those coins that have a business function) and various business cycles is far less clear than in the equity market. Perhaps when the market becomes more sophisticated and such technological understanding becomes more mainstream knowledge, this could change the landscape of cryptocurrency investing. This could lead to the development of better understood and widely disseminated cryptocurrency investment principles, which may drive more predictable investment patterns during different market cycles. Such dynamics are likely to drive further differentiation in cryptocurrency price patterns in varying market cycles and may lead to further diversification benefits as the market approaches greater levels of maturity.

No analysis is without its limitations. There are several limitations in our work. First, we have only looked at a collection of 40 cryptocurrencies. This could be extended, and include a much wider variety of cryptocurrency securities. The difficulty here is that many smaller coins do not have sufficient time windows for us to analyze. However, as time goes on, doing such analysis on a larger number of coins will become easier and may provide more robust insights. Furthermore, we could extend our portfolio sampling analysis to explicitly study diversification benefits during various market conditions. In the near future, we may be able to compare a short and intense market crisis such as the COVID-19 market crash or the Bitmex crash with that of the Russian financial crisis - or something more protracted and systemic. At present, the data is most likely insufficient.

\section{Conclusion}
\label{sec:conclusion}

Overall, we have uncovered nuanced similarities and differences between the cryptocurrency and equity markets. These mathematical properties signal increased signatures of maturity in the collective dynamics and diversification benefit of different portfolio spreads and provide concrete suggestions to retail investors seeking a relatively low-complexity exposure to the cryptocurrency market. Cryptocurrency decile sectors have been shown to share several properties, but not all, with industry sectors of the equity markets, and the most relevant findings for small-scale investors interested in limited size portfolios are shared.

There are a variety of opportunities for future research building upon the methodologies we have developed and insights highlighted in this paper. First, it would be interesting to test how the cryptocurrency market compares with other more mature and better established asset classes with respect to various statistical properties. A thorough inspection of metrics such as drawdowns, peak-to-trough analysis, changepoint propagation, intra and inter asset correlations, etc. could reveal information as to how cryptocurrencies can complement a multi-asset investment portfolio. An additional avenue of future research could be studying the data on a higher sampling rate than daily data. Given the significant composition of day traders in the cryptocurrency market, we may see patterns that deviate from what we see intra-day within the equity market. In a somewhat-related manner, studying these securities in a longer time horizon may highlight regime shifts in dynamics or optimal portfolio construction. One of the key assumptions in this work is our separating cryptocurrencies into size deciles. Further work could look into alternative bucketing criteria such as sector-allocation, volatility or other measures of risk. Finally, given the number of quantitative investment firms with burgeoning high-frequency cryptocurrency trading operations, one could examine the effectiveness of frequency-based trading strategies to see if there is greater "power" with certain trading windows. This could reveal typical holding periods for investment firms who trade in the cryptocurrency market.

\section*{Funding}
This research received no external funding.

\section*{Author contributions}
Both authors contributed equally in every aspect of the paper.

\section*{Data availability statement}
The data analyzed in this article are publicly available at \cite{Yahoo_crypto}.

\section*{Conflicts of interest}
The authors declare no conflict of interest.


\begin{adjustwidth}{-\extralength}{0cm}

\reftitle{References}
\bibliography{__NewRefsnow}


\end{adjustwidth}
\end{document}